\documentclass[12pt]{article}
\usepackage{a4,amsfonts,amsmath,bbm}

\begin{document}

\def\V#1{\mathcal{V}_{#1}}
\def\R#1{\mathcal{R}_{#1}}
\def\M#1{\mathcal{M}_{#1}}
\def\W{\mathcal{W}}
\def\Vbar#1{\overline{\mathcal{V}}_{#1}}
\def\Rbar#1{\overline{\mathcal{R}}_{#1}}
\def\Mbar#1{\overline{\mathcal{M}}_{#1}}

\def\vac{\mathit{\Omega}}
\def\logvac{\mathit{\omega}}
\def\one{\mathit{\phi}}
\def\logone{\mathit{\psi}}
\def\spin{a}
\def\index{\lambda}

\def\Nop{\hat{\mathcal{N}}}
\def\ndop{\hat{\delta}}    
\def\ndbarop{\hat{\bar{\delta}}}
\def\Lop{L}                
\def\Lbarop{\overline{L}}  
\def\Wop{W}
\def\Wbarop{\overline{W}{}}
\def\qop{\hat{q}}          
\def\hop{\hat{h}}         
\def\hbarop{\hat{\bar{h}}}
\def\Pop{\hat{\mathcal{P}}} 
\def\Pdaggerop{\Pop^{\dagger}} 
\def\Pbarop{\hat{\overline{\mathcal{P}}}}
\def\Pbardaggerop{\Pbarop{}^{\dagger}}
\def\del{\partial}
\def\Xop{{\cal O}}
\def\Xbarop{\overline{\Xop}}
\def\c{\mbox{c}}

\def\ket#1{\big |#1 \big \rangle}
\def\bra#1{\big \langle #1 \big |}
\def\braket#1#2{\big \langle #1 \big | #2\big \rangle}

\def\Rbs#1{\mathit{R}_{#1}}
\def\Vbs#1{\mathit{V}_{#1}}
\def\Xbs#1{\mathit{X}_{#1}}
\def\Ybs#1{\mathit{Y}_{#1}}
\def\Zbs#1{\mathit{Z}_{#1}}
\def\Mbs#1{\mathit{M}_{#1}}

\def\RNbs{\Rbs{0}}
\def\RObs{\Rbs{1}}
\def\VNbs{\Vbs{0}}
\def\VObs{\Vbs{1}}
\def\RNObs{\Rbs{01}}
\def\RONbs{\Rbs{10}}
\def\VOEbs{\Vbs{-1/8}}
\def\VTEbs{\Vbs{3/8}}
\def\XNbs{\Xbs{0}}
\def\XObs{\Xbs{1}}
\def\YNbs{\Ybs{0}}
\def\YObs{\Ybs{1}}
\def\bs{B}
\def\bsc{C}

\def\half{\frac{1}{2}}
\def\U{\mathcal{U}}
\def\Udagger{\U^{\dagger}}
\def\nn{{}}
\def\id{{\mathbbm 1}}
\def\ol{\overline}
\def\cross{\times}

\def\ie{i.\,e.}
\def\wrt{w.\,r.\,t.\ }
\def\etal{{\em et al.}\ }
\def\ala{{\em {\`a} la\/}}
\def\eg{e.\,g.}

\def\character#1{\chi^{\ph{\V{}}}_{#1}}

\def\e{\mathrm{e}}
\def\imag{\mathrm{i}}

\def\ph#1{\phantom{#1}}
\def\phn{{\ph{0}}}
\def\phm{{\ph{-}}}

\def\beqn{\begin{eqnarray}}
\def\eeqn{\end{eqnarray}}

\def\binomial#1#2{\left(\begin{array}{c}{#1}\\{#2}\end{array}\right)}
\def\fused{\otimes}
\def\fuse#1#2{\big(#1\fused #2\big)}
\def\fuseprod#1#2{\left(#1\fused_{\text{\footnotesize f}} #2\right)}
\def\abs#1{\left|#1\right|}
\def\gsu{\hat{g}}
\def\fsu{\hat{f}}
\def\chibar{\ol{\chi}}

\def\bb#1{\mathord{\hbox{\boldmath$#1$}}}

\def\switchtitlefootnote{\renewcommand{\thefootnote}{\fnsymbol{footnote}}}
\def\switchtextfootnote{\renewcommand{\thefootnote}{\arabic{footnote}}}

\def\abstracttext{We investigate the set of boundary states in the symplectic
fermion description of the logarithmic conformal field theory with central
charge $\c=-2$. We show that the thus constructed states correspond exactly to 
those derived under the restrictions of the maximal chiral symmetry algebra for 
this model, $\W(2,3,3,3)$. This connects our previous work to the coherent 
state approach of Kawai and Wheater.}

\switchtitlefootnote
\pagestyle{empty}
\begin{titlepage}

\begin{center}

\vspace*{-1.0cm}
{\hbox to \hsize{\hfill hep-th/0207181}}
{\hbox to \hsize{\hfill ITP-UH-18/02}}
\vspace*{1.0cm}

\vspace*{1.5cm}
{\Large \bf 
        Boundary States and\\
	Symplectic Fermions}

\vspace*{1.5cm}
{       Andreas Bredthauer\footnote[1]{\parbox[t]{12cm}{\tt 
        Andreas.Bredthauer@itp.uni-hannover.de}}}\\[1cm]

{\em	
        Institut f{\"u}r Theoretische Physik\\
        Universit{\"a}t Hannover\\
        Appelstra{\ss}e 2, 30167 Hannover, Germany}

\vspace*{1cm}
{\small July 19, 2002}

\vspace*{5cm}
{\small \bf Abstract}\\[0.5cm]

\parbox{12.5cm}
{\small
\hspace*{0.7cm}\abstracttext
}
\end{center}
\end{titlepage}

\newpage

\setcounter{page}{1}
\switchtextfootnote
\pagestyle{plain}

\section{Introduction}
During the last 20 years, conformal field theory (CFT) in two dimensions 
\cite{BPZ:84} has become a very important tool in theoretical physics. 
Especially, two different directions are subject of current interest: The 
study of critical systems on surfaces involving boundaries led to a good 
knowledge of the so-called boundary CFTs (BCFT) 
\cite{Is:89,Ca:89,BePePeZu:98,BePePeZu:00}. On the other hand, already in 1991, 
Saleur showed the existence of density fields with scaling dimension zero 
occurring in the treatment of dense polymers  \cite{Sa:92}. These fields may 
cause the existence of operators yielding logarithmically diverging correlation 
functions. The two subjects, BCFT and logarithmic conformal field theory 
(LCFT), enjoy increasing popularity in both condensed matter physics and string 
theory. 

Even though there has been much progress in the field, LCFTs are not yet
completely understood. However, it has been found that many properties of
ordinary rational CFTs can be generalized to LCFT, such as characters, partition
functions and fusion rules, see, \eg,
\cite{Fl:96,GaKa:96,GaKa:961,Ro:96,Fl:97,GaKa:98,FjFuHwSeTi:02,Mi:01} and
\cite{Ga:01,Fl:01} for some recent reviews. In ordinary CFTs, especially in 
unitary minimal models, the presence of a boundary is mathematically and 
physically described by a standard procedure introduced by Ishibashi 
\cite{Is:89} and Cardy \cite{Ca:89} that allows to derive boundary states 
encoding the physical boundary conditions. Unfortunately, LCFTs involving a 
boundary happen to be more difficult to treat. There have been different 
approaches towards a consistent description of boundary LCFTs in terms of 
boundary states emerging first for two years ago
\cite{KoWh:00,Ism:01,KaWh:01,BrFl:02}, see also \cite{Ka:02,Ka:021}. LCFT in 
the vicinity of a boundary is also dealt with in \cite{MoRo:00,Le:00}. All those 
works focus on the best understood example of a LCFT, the $\c=-2$ realization 
with the maximally extended chiral symmetry algebra $\W(2,3,3,3)$. The earlier
results are different and partly contradictory. Most successful seem the 
ideas of Kawai and Wheater \cite{KaWh:01} using symplectic fermions and coherent
states and of ourselves \cite{BrFl:02}. The concept of symplectic fermions was 
first introduced by Kausch \cite{Ka:00} in order to describe the rational 
$\c=-2$ (bulk) LCFT. In our own work, a general, very basic approach towards the
derivation of boundary states in the case of the $\W$-algebra is presented that 
allows to handle complicated structures such as indecomposable representations 
in LCFTs.

This letter is positioned exactly at this point. We show that the two different
symmetries -- the symplectic fermions vs.\ $\W(2,3,3,3)$-algebra -- lead to the 
same set of boundary states. In particular, the former one, though extending 
the latter, implies no additional restrictions on the boundary states. By this, 
we can show that the coherent state approach is fully equivalent to ours 
yielding the same results. This corresponds to the presumption of Kawai 
\cite{Ka:021} that the coherent states are indeed as good as taking the usual 
Ishibashi states.

The paper proceeds as follows: 
In section 2, a short introduction to the rational $\c=-2$ LCFT is given both
for the $\W(2,3,3,3)$-algebra and in the symplectic fermion picture. 
Then, section 3 and 4 review the results of Kawai and Wheater 
and those deduced by us. 
In section 5, the boundary states for the symplectic fermion symmetry algebra 
are derived using the method of \cite{BrFl:02} and compared to both of the 
previous results. Finally, section 6 concludes the paper with a short 
discussion. 

\section{The model}

The CFT realization at $\c=-2$ is based on the extended chiral symmetry algebra 
$\W(2,3,3,3)$ consisting of the energy-momentum tensor $\Lop(z)$ and a triplet 
of spin-3 fields $\Wop^\spin(z)$. With the two quasi-primary normal-ordered 
fields 
$\Lambda=\;:\!\!\Lop^2\!\!:\!-3/10\,\del^2\Lop$ and 
$V^{\spin}=\;:\!\!\Lop\Wop^{\spin}\!\!:\!-3/14\,\del^2\Wop^{\spin}$ the 
commutation relations for the corresponding modes read:
\begin{align}
  \big[\Lop_m,\Lop_n\big]\,=&(m-n)\Lop_{m+n}-\frac{1}{6}
    \left(m^3-m\right)\delta_{m+n,0}, \nonumber \\
  \big[\Lop_m^{\ph{r}},\Wop_n^a\big]=&(2m-n)\Wop_{m+n}^a, \nonumber \\
  \big[\Wop_m^a,\Wop_n^b\big]=
    &\gsu^{\,ab}\left(2\left(m-n\right)\Lambda_{m+n}
     +\frac{1}{20}\left(m-n\right)
     \left(2m^2+2n^2-mn-8\right)\Lop_{m+n}\right. \nonumber \\
    &\ph{\gsu^{\,ab}\Big(}\left.-\frac{1}{120}m\left(m^2-1\right)
     \left(m^2-4\right)\delta_{m+n,0}\right) 
     \label{eqn:commutation-relations} \\
    &\vspace*{1cm}+\fsu^{\,ab}_c\left(\frac{5}{14}\left(2m^2+2n^2-3mn-4\right)
     \Wop^c_{m+n}+\frac{12}{5}V^c_{m+n}\right). \nonumber
\end{align}
Here, $\gsu^{\,ab}$ is the metric and $\fsu^{\,ab}_c$ are the structure 
constants of $su(2)$. It is convenient to arrange the fields in a Cartan-Weyl 
basis $\Wop^0$, $\Wop^{\pm}$. In this framework, we have $\gsu^{\,00}=1$, 
$\gsu^{\,+-}=\gsu^{\,-+}=2$, $\fsu^{\,0\pm}_{\pm}=-\fsu^{\,\pm 0}_{\pm}=\pm 1$,
and $\fsu^{\,+-}_0=-\fsu^{\,-+}_0=2$.

The algebra yields a set of six representations that close under fusion. There
are four ordinary highest weight representations: $\V{0}$ is based on the vacuum
state $\vac$ with weight $h=0$, $\V{-1/8}$ emerges from the state $\mu$ with
weight $h=-1/8$. Then, one has two doublet representations $\V{1}$ based on the 
states $\one^\pm$ and $\V{3/8}$ built from $\nu^\pm$. Furthermore, two 
{\em indecomposable\/} or {\em generalized highest weight} representations 
$\R{0}$ and $\R{1}$ emerge. They base on the states $\logvac$ and 
$\logone^\pm$, respectively. These states form rank-2 Jordan blocks in $\Lop_0$ 
together with the states $\vac$ and $\one^\pm$. Thus, $\V{0}$ and $\V{1}$ are 
subrepresentations of $\R{0}$ and $\R{1}$, respectively. $\R{0}$ also contains 
two subrepresentations of type $\V{1}$ built on the states
\begin{eqnarray} 
 \begin{array}{rcl@{\qquad}rcl}
  \Psi_1^+&\!\!\!=\!\!\!&\Wop_{-1}^+\logvac,&
  \Psi_2^+&\!\!\!=\!\!\!&\left(\Wop_{-1}^0+\half\Lop_{-1}^{\phn}\right)\logvac, 
  \vspace{1ex} \\
  \Psi_1^-&\!\!\!=\!\!\!&\left(-\Wop_{-1}^0+\half\Lop_{-1}^{\phn}\right)\logvac,
 &\Psi_2^-&\!\!\!=\!\!\!&\Wop_{-1}^-\logvac.
 \end{array} 
\end{eqnarray}
For the bulk states of the $\R{0}$ and $\R{1}$ we use the metric of 
\cite{BrFl:02} that reads:
\beqn
  \begin{array}{ccl@{\qquad}ccl@{\qquad}ccl}
    \braket{\vac}{\vac}&\!\!\!=\!\!\!&0,&
    \braket{\vac}{\logvac}&\!\!\!=\!\!\!&1,&
    \braket{\logvac}{\logvac}&\!\!\!=\!\!\!&d, \vspace{1ex} \\
    \braket{\one^+}{\one^-}&\!\!\!=\!\!\!&0,&
    \braket{\one^+}{\logone^-}&\!\!\!=\!\!\!&-1,&
    \braket{\logone^+}{\logone^-}&\!\!\!=\!\!\!&-t,
  \end{array}
  \label{eqn:metricR}
\eeqn  
where $d$ and $t$ are in principle arbitrary real numbers. The fusion rules for 
this model read:
\beqn
  \begin{array}{rcl@{\qquad}rcl@{\qquad}rcl}
  \V{0}\cross\Psi&\!\!\!=\!\!\!&\Psi,& 
   \V{-1/8}\cross\V{-1/8}&\!\!\!=\!\!\!&\R{0},&  
    \V{3/8}\cross\V{1}&\!\!\!=\!\!\!&\V{-1/8}, \vspace*{1ex}\\
  \V{1}\cross\V{1}&\!\!\!=\!\!\!&\V{0},&
   \V{-1/8}\cross\V{3/8}&\!\!\!=\!\!\!&\R{1},&  
    \V{-1/8}\cross\R{m}&\!\!\!=\!\!\!&2\,\V{-1/8}+2\,\V{3/8}, \vspace*{1ex}\\
  \V{1}\cross\R{0}&\!\!\!=\!\!\!&\R{1},&
   \V{-1/8}\cross\V{1}&\!\!\!=\!\!\!&\V{3/8},&  
    \V{3/8}\cross\R{m}&\!\!\!=\!\!\!&2\,\V{-1/8}+2\,\V{3/8}, \vspace*{1ex}\\
  \V{1}\cross\R{1}&\!\!\!=\!\!\!&\R{0},&
   \V{3/8}\cross\V{3/8}&\!\!\!=\!\!\!&\R{0},&  
    \R{m}\cross\R{n}&\!\!\!=\!\!\!&2\,\R{0}+2\,\R{1}.
  \end{array}  
  \label{eqn:fusion-rules}  
\eeqn        
Here, $m$ and $n$ can take the values $0,\,1$. From \eqref{eqn:fusion-rules} one
reads off that $\big\{\R{0},\,\R{1},\,\V{-1/8},\,\V{3/8}\big\}$ is a sub-group 
closed under fusion itself. The characters for the model are given by:
\begin{align}
  \character{\V{0}}(q)&=\frac{1}{2\eta(q)}\left(
    \Theta_{1,2}^{\phn}(q)+\left(\del\Theta\right)\!{}_{1,2}^{\phn}(q)\right), 
  &\character{\V{-1/8}}(q)&=\frac{1}{\eta(q)}
    \Theta_{0,2}^{\phn}(q),\nonumber \\
  \character{\V{1}}(q)&=\frac{1}{2\eta(q)}\left(
    \Theta_{1,2}^{\phn}(q)-\left(\del\Theta\right)\!{}_{1,2}^{\phn}(q)\right), 
  &\character{\V{3/8}}(q)&=\frac{1}{\eta(q)}
    \Theta_{2,2}^{\phn}(q),\label{eqn:triplet_characters}\\
  \character{\R{}}(q)&\equiv\character{\R{0}}(q)
    =\character{\R{1}}(q)=\frac{2}{\eta(q)}\Theta_{1,2}^{\phn}(q).
    \hspace*{-2cm}\nonumber
\end{align}
Note, that the physical characters are only $\character{\V{-1/8}}$,
$\character{\V{3/8}}$ and $\character{\R{}}$ forming a three-dimen\-sio\-nal 
representation of the modular group that corresponds to the above-mentioned 
sub-group. Here, $\eta(q)=q^{1/24}\prod_{n\in N}\left(1-q^n\right)$ is the 
Dedekind eta function and $\Theta_{r,2}(q)$ and 
$\left(\del\Theta\right)_{1,2}\!(q)$ are the ordinary and affine 
Riemann-Jacobi theta functions:
\beqn
    \Theta_{r,k}(q)=\sum_{n\in {\mathbbm Z}}q^{\left(2kn+r\right)^2/4k}
    \mbox{~~and~~}
  \left(\del\Theta\right)_{r,k}\!(q)=
    \sum_{n\in {\mathbbm Z}}\left(2kn+r\right)q^{\left(2kn+r\right)^2/4k}.
  \label{eqn:theta-functions}
\eeqn
In ordinary CFTs, the characters coincide with the torus amplitudes. Here, this
is no longer the case: The torus amplitudes form a slightly larger,
five-dimensional representation of the modular group. It reads:
\beqn
  \character{\V{0}},~\character{\V{-1/8}},~\character{\V{1}},~
  \character{\V{3/8}},\mbox{~and~}
  \character{\widetilde{\R{}}}(q)\equiv\frac{2}{\eta(q)}
    \left(\Theta_{1,2}^{\phn}(q)+
          \imag\alpha\log(q)\left(\del\Theta\right)_{1,2}\!(q)
    \right).
  \label{eqn:torus-amplitudes}  
\eeqn
This representation was analyzed by Flohr \cite{Fl:96}. There, the 
${\cal S}$-matrix transforming the ``characters'' under 
$\tau\longrightarrow-1/\tau$ was constructed and it was shown that it yields 
the fusion rules \eqref{eqn:fusion-rules} only in the limit 
$\alpha\longrightarrow 0$ under which the logarithmic term in 
\eqref{eqn:torus-amplitudes} vanishes. However, in this limit, {\cal S} became 
singular.
      
There exists an explicit Lagrangian formulation for the $\c=-2$ LCFT based on 
two fermionic fields $\eta$ and $\xi$ of scaling dimension $1$ and $0$, 
respectively:
\beqn
  S=\frac{1}{\pi}\int
  \mbox{d}^2z\left(\eta\bar\del\xi+\bar\eta\del\bar\xi\right).
\eeqn
This is the fermionic ghost system at $\c=-2$ with the operator product
expansions
\beqn
  \eta(z)\xi(w)=\xi(z)\eta(w)=\frac{1}{z-w}+\ldots\,.
\eeqn
All other products are regular. Kausch \cite{Ka:00} showed, that these two 
fields combine into a two-component symplectic fermion
\beqn
  \chi^+\equiv\eta\mbox{~~and~~}\chi^-\equiv\del\xi.
\eeqn
The choice assures that $\chi^+$ and $\chi^-$ have the same conformal weight 
$h=1$. This description differs from the ghost system only by the treatment of 
the zero modes in $\chi^-$ and $\xi$. The fermion modes are defined by the 
usual power series expansion
\beqn
  \chi^\pm(z)=\sum_{m\in\mathbbm{Z}+\lambda}\chi^\pm_m z^{-m-1},
\eeqn
where $\lambda=0$ in the untwisted (bosonic) sector and $\lambda=\half$ in 
the twisted (fermionic) sector. The modes satisfy the anticommutation relations
\beqn
  \big\{\chi^{\,\alpha}_m,\,\chi^{\,\beta}_n\big\}=
  m\,\varepsilon^{\alpha\beta}\delta_{m+n,0}\,,
\eeqn
with the totally antisymmetric tensor $\varepsilon^{\pm\mp}=\pm 1$. The 
symplectic fermion decomposes the Virasoro modes and the modes of the three
spin-3 fields $\Wop(z)$ of $\W(2,3,3,3)$ \cite{GaKa:96,GaKa:961,Ga:01}:
\begin{align}
 \begin{split}
  \Lop_n&=\half \varepsilon_{\alpha\beta}\sum_{j\in\mathbbm{Z}+\lambda}
  :\chi^{\,\alpha}_j \chi^{\,\beta}_{n-j} :+\frac{\lambda(\lambda-1)}{2}
  \delta_{n,0}, \\
  \Wop^0_n&=-\half\sum_{j\in\mathbbm{Z}+\lambda} j\cdot\big\{:\chi^+_{n-j}
  \chi^-_j:+:\chi^-_{n-j}\chi^+_j:\big\}, \\
  \Wop^\pm_n&=\sum_{j\in\mathbbm{Z}+\lambda} j\cdot \chi^{\pm}_{n-j}
  \chi^{\pm}_j.
 \end{split}
 \label{eqn:chi--W}
\end{align}
The highest weight states become related to each other by introducing the 
fermion symmetry: In the twisted sector, the doublet states of weight $h=3/8$
are connected to the singlet at weight $h=-1/8$ by 
$\nu^\alpha=\chi^\alpha_{-1/2}\mu$. The states of weight $0$ in the untwisted 
sector are related by $\xi^\pm=-\chi^\pm_0\logvac$, 
$\vac=\chi^-_0\chi^+_0\logvac$. Furthermore, one finds 
$\one^\alpha=\chi^\alpha_{-1}\vac$ and $\logone^\alpha=\chi^\alpha_{-1}\logvac$.
Thus, this additional symmetry intertwines the representation $\R{0}$ with 
$\R{1}$ and $\V{-1/8}$ with $\V{3/8}$.

\section{Approach 1: Coherent boundary states 
\label{section:sympl_coherent_states}}

Starting point for any derivation of boundary conditions is the absence of
energy-mo\-men\-tum flow across the boundary and corresponding gluing 
conditions for the extended symmetry fields. On a cylinder, the boundary 
conditions are identified with an initial and final state of a propagating 
closed string: the {\em boundary states} $\ket{\bs}$. After radial ordering and 
in the framework of symplectic fermions this yields the following consistency 
equations:
\beqn
  &\left(\Lop_n-\Lbarop_{-n}\right)\ket{\bs}=0, \label{eqn:VirB} \\[1ex]
  &\left(\chi^\pm_n-\e^{\pm\imag\phi}\overline{\chi}{}^\pm_{-n}\right)
  \ket{\bs}=0, 
  \label{eqn:chiB}
\eeqn
where $\phi$ is a phase that occurs in the gluing condition of $\chi$ and
$\ol{\chi}$. The latter equation implies the first one due to 
\eqref{eqn:chi--W}. Kawai and Wheater showed that \eqref{eqn:chiB} is solved 
by the coherent states \cite{KaWh:01}
\beqn
  \ket{B_{0\phi}}=N\exp\left(\sum_{k>0}\frac{\e^{\imag\phi}}{k}\chi^-_{-k}
  \ol{\chi}{}^+_{-k}+\frac{\e^{-\imag\phi}}{k}\ol{\chi}{}^-_{-k}\chi^+_{-k}
  \right)
  \ket{0_\phi}.
\eeqn
Here, $N$ is a normalization factor and $\ket{0_\phi}$ is a non-chiral ground
state. The boundary states were designed in such a way that they are compatible
with the $\W$-algebra and thus obey \eqref{eqn:VirB} and
\beqn
  \left(\Wop^\spin_n+\Wbarop^\spin_{-n}\right)\ket{\bs}=0.
  \label{eqn:WB}
\eeqn  
This implies that the phase $\phi$ can only take the values $\phi=0$ and 
$\phi=\pi$. Therefore, the non-chiral ground states are given by the 
``invariant vacua'' $\big\{\fuse{\vac}{\ol{\vac}},\,
\fuse{\logvac}{\ol{\logvac}},\,\fuse{\mu}{\ol{\mu}}\big\}$. This yields six 
possible boundary states, denoted by $(+)$ if $\phi=0$ and $(-)$ for 
$\phi=\pi$:
\beqn
  \ket{B_{\vac+}}\equiv\ket{B_{\vac,\phi=0}},~\ket{B_{\vac-}},
 ~\ket{B_{\logvac\pm}},\mbox{~and~}\ket{B_{\mu\pm}}.
\eeqn
The corresponding cylinder amplitudes are given by the natural pairings
$\bra{\bs}\qop\ket{\bsc}=\bra{\bs}q^{\cal H}\ket{\bsc}=
\bra{\bs}(q^{1/2})^{(\Lop_0+\Lbarop_0+1/6)}\ket{\bsc}$. For the interesting 
(untwisted) sector, they are 
\beqn
  \bordermatrix{&\ket{B_{\vac+}}&\ket{B_{\vac-}}&\ket{B_{\logvac+}}&
    \bra{B_{\logvac-}} \vspace*{1ex}\cr
    \bra{B_{\vac+}}&0&0&\eta(q)^2&\Theta_{1,2}(q) \vspace*{1ex} \cr
    \bra{B_{\vac-}}&0&0&\Theta_{1,2}(q)&\eta(q)^2 \vspace*{1ex} \cr
    \bra{B_{\logvac+}}&\eta(q)^2&\Theta_{1,2}(q)&d(d+\ln(q))\eta(q)^2&
      d(d+\ln(q))\Theta_{1,2}(q) \vspace*{1ex} \cr
    \bra{B_{\logvac-}}&\Theta_{1,2}(q)&\eta(q)^2&
      d(d+\ln(q))\Theta_{1,2}(q)&d(d+\ln(q))\eta(q)^2 \vspace*{1ex}
  }.
\eeqn
The different factors and signs in contrast to \cite{KaWh:01} arise due to our 
different normalization of the metric. To get rid of the unphysical terms 
proportional to $\log(q)\Theta_{1,2}\!\,(q)$, one of the states 
$\ket{B_{\logvac\pm}}$ was discarded and the physical boundary conditions were 
derived with this reduced set. This was possible according to the ${\mathbbm
Z}_2$ symmetry $\phi\longrightarrow\phi+\pi\mod 2\pi$. 

Candidates for the Ishibashi states were deduced by diagonalizing the cylinder 
amplitudes, \ie, $\bra{i}\qop\ket{j}=\delta^{\phn}_{ij}\character{i}\!(q)$. 
However, it was not possible to express the physical boundary states in terms 
of this basis. Kawai and Wheater proposed the following five states and five 
corresponding duals:
\beqn
  \begin{array}{rcl@{\qquad}rcl}
    \ket{\VNbs}&\!\!\!=\!\!\!&\half\ket{B_{\vac+}}+\half\ket{B_{\vac-}},&
    \bra{\VNbs}&\!\!\!=\!\!\!&-\half\bra{B_{\logvac+}}-\half\bra{B_{\logvac-}},
    \vspace{2ex} \\
    \ket{\VObs}&\!\!\!=\!\!\!&\half\ket{B_{\vac+}}-\half\ket{B_{\vac-}},&
    \bra{\VObs}&\!\!\!=\!\!\!&\half\bra{B_{\logvac+}}-\half\bra{B_{\logvac-}},
    \vspace{2ex} \\
    \ket{\VOEbs}&\!\!\!=\!\!\!&\half\ket{B_{\mu+}}+\half\ket{B_{\mu-}},&
    \bra{\VOEbs}&\!\!\!=\!\!\!&\half\bra{B_{\mu+}}+\half\bra{B_{\mu-}},
    \vspace{2ex} \\
    \ket{\VTEbs}&\!\!\!=\!\!\!&\half\ket{B_{\mu+}}-\half\ket{B_{\mu-}},&
    \bra{\VTEbs}&\!\!\!=\!\!\!&\half\bra{B_{\mu+}}-\half\bra{B_{\mu-}},
    \vspace{2ex} \\
    \ket{\Rbs{\,}}&\!\!\!=\!\!\!&\sqrt{2}\,\ket{B_{\vac+}},&
    \bra{\Rbs{\,}}&\!\!\!=\!\!\!&-\sqrt{2}\,\bra{B_{\logvac-}}.
  \end{array}
  \label{eqn:KaWh-IBS}
\eeqn
The (ket-)states form only a four-dimensional space. Especially,
$\ket{\Rbs{\,}}$ is associated to the indecomposable representations but only
built on the subrepresentations. It is evident that the states 
$\ket{B_{\logvac\pm}}$ cannot obey equation \eqref{eqn:chiB} without further 
restrictions because they are based on the state $\fuse{\logvac}{\ol{\logvac}}$ 
which is obviously not a proper ground state:
\beqn
  \big[\Lop_0-\Lbarop_0\big]\fuse{\logvac}{\ol{\logvac}}=\fuse{\vac}{\ol{\logvac}}-
  \fuse{\logvac}{\ol{\vac}}\neq 0,
\eeqn 
unless the right-hand side state is discarded as in the unique local $\c=-2$
LCFT \cite{GaKa:98}. There, a chiral and an anti-chiral version of the rational
$\c=-2$ LCFT are glued together to obtain a non-chiral theory. In order to keep
locality of the correlators, certain states had to be divided out, namely the
image of $(\Lop_0-\Lbarop_0)$. This was not mentioned by Kawai and Wheater. It 
is shown in the following that their considerations are indeed compatible with 
the result of \cite{BrFl:02} and lead to the same results if starting from the 
``vacua'' of the complete chiral theory.

\section{Approach 2: Boundary states for the $\W$-algebra \label{section:W-BS}}

In \cite{BrFl:02} the span of boundary states under the constraints of the 
$\W(2,3,3,3)$-algebra was derived. This was done by inventing a 
straight-forward method that uses only basic properties of the theory and its 
representations. Due to that it was possible to keep especially the inner 
structure of the indecomposable representations $\R{0}$ and $\R{1}$ and their 
subrepresentations visible. This allowed to find relations between the derived 
states. Ten boundary states were identified: 

The states $\ket{\VOEbs}$ and $\ket{\VTEbs}$ corresponding to the admissible
irreducible representations $\V{-1/8}$ and $\V{3/8}$ are the usual Ishibashi 
states for these modules.

For the indecomposable representations $\R{\index}$, to stay close to the usual 
notions, the definition of the Ishibashi states was generalized. The two 
states 
\beqn
  \ket{\Rbs{\index}}=\sum_{l,m,n}\gamma^{\index\,l\,m}_{\ph{\index\,l\,m}n}
    \fuse{\id}{\ol{\U}}\ket{l,m}\otimes\ol{\ket{l,n}},~~\lambda=0,\,1
\eeqn
are called {\em generalized\/} Ishibashi states. Here,
$\big\{\ket{l,m};\,l=h,h+1,\ldots,~m=1,\ldots\big\}$ is an arbitrary basis over 
the representation $\R{\index}$ where $l$ counts the levels beginning from the 
top-most, which is $h=0$ in our case. The basis states on each level of the 
representation are counted by $m$. Similarly, $\big\{\ol{\ket{l,n}}\big\}$ is 
the basis for the anti-holomorphic module $\ol{\R{\index}}$. The matrix 
$\gamma^\index$ was identified to be the inverse metric on $\R{\index}$. In 
ordinary CFTs, these bases can be chosen orthonormal and then the result would 
coincide with the usual Ishibashi state. It was argued in \cite{BrFl:02} that 
this is not applicable here.

The Ishibashi states corresponding to the two subrepresentations $\V{0}$ and
$\V{1}$ were derived with the help of an operator $\Nop=\ndop+\ndbarop$, where
$\ndop$ is the off-diagonal part of $\Lop_0$ that was considered to be in Jordan
form. Since there are rank-2 Jordan cells at most, $\ndop^2=0$ and thus,
$\Nop^3=0$. It was argued that the states
\beqn
  \ket{\Vbs{\index}}=\half\Nop\ket{\Rbs{\index}}
\eeqn
do not vanish and fulfill \eqref{eqn:VirB} and \eqref{eqn:WB}, \ie, are properly
defined boundary states. These are called {\em level-2\/} Ishibashi states and 
contain only contributions from the corresponding subrepresentations.

In addition, two doublets of states were found that glue together the two
different indecomposable representations $\R{0}$ and $\R{1}$ at the boundary. 
They were given in terms of operators $\Pop$ and $\Pdaggerop$ that intertwine 
the two representations and have the following action on the (bulk) states:
\beqn
  \begin{array}{rcl@{\qquad}rcl@{\qquad}rcl}
    \Pdaggerop_\pm\ket{\logvac}&\!\!\!=\!\!\!&\ket{\xi^\pm},&
    \Pdaggerop_\pm\ket{\vac}&\!\!\!=\!\!\!&0,& 
    \Pop_+\ket{\logone^\pm}&\!\!\!=\!\!\!&-\ket{\Psi_2^\pm},
    \vspace{1.5ex} \\
    \Pop_\pm\ket{\xi^\mp}&\!\!\!=\!\!\!&\pm\ket{\vac},&
    \Pop_\pm\ket{\one^\pm}&\!\!\!=\!\!\!&0,& 
    \Pop_-\ket{\logone^\pm}&\!\!\!=\!\!\!&\ket{\Psi_1^\pm}.
  \end{array}
  \label{eqn:P}
\eeqn
This yields the so-called {\em mixed\/} Ishibashi states $\ket{\RNObs^\pm}$ and
$\ket{\RONbs^\pm}$:
\beqn
  \begin{array}{ccccc@{\qquad}ccccc}
    \ket{\RNObs^\pm}&\!\!\!=\!\!\!&\Pop_\pm\ket{\RObs}
                    &\!\!\!=\!\!\!&\Pbardaggerop_\pm\ket{\RNbs}, 
   &\ket{\VNbs}&\!\!\!=\!\!\!&\Pbarop_\mp\ket{\RNObs^\pm}
               &\!\!\!=\!\!\!&\Pop_\mp\ket{\RONbs^\pm}, \vspace{1ex} \\
    \ket{\RONbs^\pm}&\!\!\!=\!\!\!&\Pbarop_\pm\ket{\RObs}
                    &\!\!\!=\!\!\!&\Pdaggerop_\pm\ket{\RNbs},
   &\ket{\VObs}&\!\!\!=\!\!\!&\Pdaggerop_\mp\ket{\RNObs^\pm}
               &\!\!\!=\!\!\!&\Pbardaggerop_\mp\ket{\RONbs^\pm}.
  \end{array}
  \label{eqn:R01}
\eeqn
These relations can be drawn schematically. It is not quite unexpected that
there is a one-to-one correspondence to the embedding scheme of the local 
theory \cite{GaKa:98}: The states that are divided out there are due to 
\eqref{eqn:VirB} exactly those that do not contribute to the boundary states.
\begin{displaymath}
 \begin{array}{l@{\qquad\qquad}r}
  \begin{picture}(140,210)(-92,-50)
    \put(2,40){\vbox to 0pt
        {\vss\hbox to 0pt{\hss$\bullet$\hss}\vss}}
    \put(2,90){\vbox to 0pt
        {\vss\hbox to 0pt{\hss$\bullet$\hss}\vss}}
    \put(62,130){\vbox to 0pt
        {\vss\hbox to 0pt{\hss$\bullet$\hss}\vss}}
    \put(62,0){\vbox to 0pt
        {\vss\hbox to 0pt{\hss$\bullet$\hss}\vss}}
    \put(56,4){\vector(-3,2){48}}
    \put(58,6){\vector(-2,3){52}}
    \put(56,126){\vector(-3,-2){48}}
    \put(58,124){\vector(-2,-3){52}}
    \put(-2,40){\vbox to 0pt
        {\vss\hbox to 0pt{\hss$\bullet$\hss}\vss}}
    \put(-2,90){\vbox to 0pt
        {\vss\hbox to 0pt{\hss$\bullet$\hss}\vss}}
    \put(-62,130){\vbox to 0pt
        {\vss\hbox to 0pt{\hss$\bullet$\hss}\vss}}
    \put(-62,0){\vbox to 0pt
        {\vss\hbox to 0pt{\hss$\bullet$\hss}\vss}}
    \put(-8,36){\vector(-3,-2){48}}
    \put(-6,84){\vector(-2,-3){52}}
    \put(-8,94){\vector(-3,2){48}}
    \put(-6,46){\vector(-2,3){52}}
    \put(55,0){\vector(-1,0){110}}
    \put(55,130){\vector(-1,0){110}}
    \put(-67,-15){\hbox to 0pt{\hss$\ket{\VNbs}$}}
    \put(67,-15){\hbox to 0pt{$\ket{\RNbs}$\hss}}
    \put(-67,135){\hbox to 0pt{\hss$\ket{\VObs}$}}
    \put(67,135){\hbox to 0pt{$\ket{\RObs}$\hss}}
    \put(-12,18){\hbox to 0pt{$\ket{\RNObs^\pm}$\hss}}
    \put(-12,107){\hbox to 0pt{$\ket{\RONbs^\pm}$\hss}}
    \put(0,-35){\hbox to 0pt{\hss{\em boundary states}\hss}}
  \end{picture}
 &
  \begin{picture}(170,210)(-92,-50)
    \put(2,40){\vbox to 0pt
        {\vss\hbox to 0pt{\hss$\bullet$\hss}\vss}}
    \put(2,90){\vbox to 0pt
        {\vss\hbox to 0pt{\hss$\bullet$\hss}\vss}}
    \put(62,130){\vbox to 0pt
        {\vss\hbox to 0pt{\hss$\bullet$\hss}\vss}}
    \put(62,0){\vbox to 0pt
        {\vss\hbox to 0pt{\hss$\bullet$\hss}\vss}}
    \put(56,4){\vector(-3,2){48}}
    \put(58,6){\vector(-2,3){52}}
    \put(56,126){\vector(-3,-2){48}}
    \put(58,124){\vector(-2,-3){52}}
    \put(-2,40){\vbox to 0pt
        {\vss\hbox to 0pt{\hss$\bullet$\hss}\vss}}
    \put(-2,90){\vbox to 0pt
        {\vss\hbox to 0pt{\hss$\bullet$\hss}\vss}}
    \put(-62,130){\vbox to 0pt
        {\vss\hbox to 0pt{\hss$\bullet$\hss}\vss}}
    \put(-62,0){\vbox to 0pt
        {\vss\hbox to 0pt{\hss$\bullet$\hss}\vss}}
    \put(-8,36){\vector(-3,-2){48}}
    \put(-6,84){\vector(-2,-3){52}}
    \put(-8,94){\vector(-3,2){48}}
    \put(-6,46){\vector(-2,3){52}}
    \put(55,0){\vector(-1,0){110}}
    \put(55,130){\vector(-1,0){110}}
    \put(-67,-15){\hbox to 0pt{\hss$\vac$}}
    \put(67,-15){\hbox to 0pt{$\logvac$\hss}}
    \put(-67,135){\hbox to 0pt{\hss$\one$}}
    \put(67,135){\hbox to 0pt{$\logone$\hss}}
    \put(-7,25){\hbox to 0pt{$\rho^{\,\pm}$\hss}}
    \put(-7,100){\hbox to 0pt{$\bar\rho^{\,\pm}$\hss}}
    \put(0,-35){\hbox to 0pt{\hss$\R{}$\hss}}
  \end{picture}
 \end{array}
\end{displaymath}
\refstepcounter{figure} \label{fig:R01}
\vspace*{-6ex}
\begin{center} {\em figure \ref{fig:R01}: boundary states vs.\ local theory}
\end{center}
The lines in the left picture in figure \ref{fig:R01} refer to the action of 
$\Nop$, $\Pop$, $\Pdaggerop$, $\Pbarop$, and $\Pbardaggerop$ while in the right 
one they denote the action of the (non-chiral) symmetry algebra.

The non-vanishing natural pairings of the boundary states are given by
\beqn
 \begin{array}{ccl@{\qquad}ccl}
  \bra{\VOEbs}\qop\ket{\VOEbs}&\!\!\!=\!\!\!&\character{\V{-1/8}}(q),&
  \bra{\VTEbs}\qop\ket{\VTEbs}&\!\!\!=\!\!\!&\character{\V{3/8}}(q),
  \vspace{2ex} \\
  \bra{\RNbs}\qop\ket{\RNbs}&\!\!\!=\!\!\!&\character{\R{}}(q),&
  \bra{\RObs}\qop\ket{\RObs}&\!\!\!=\!\!\!&\character{\R{}}(q).
 \end{array}
 \label{eqn:bspairing}
\eeqn
These coincide with the physical characters forming the three-dimensional
representation of the modular group. The torus amplitudes, on the other hand, 
are first seen with the help of additional, so-called {\em weak\/} boundary 
states $\ket{\Xbs{\index}}$ and $\ket{\Ybs{\index}}$, $\index=0,1$, that obey
\beqn
  \ket{\Rbs{\lambda}}=\Nop\ket{\Xbs{\lambda}}+\ket{\Ybs{\lambda}}.
\eeqn
These states could be chosen uniquely in such a way that they serve as the duals
to the null states $\ket{\Vbs{\index}}$ obtaining
\beqn
  \begin{array}{rcl@{\qquad}rcl}
    \bra{\Xbs{\index}}\qop\ket{\Vbs{\index}}&\!\!\!=\!\!\!&
    \character{\V{\index}}(q),&
    \bra{\Xbs{\index}}\qop\ket{\Rbs{\index}}&\!\!\!=\!\!\!&
    \log(q)\cdot\character{\V{\index}}(q), \vspace{1.5ex} \\
    \bra{\Xbs{\index}}\qop\ket{\Ybs{\index}}&\!\!\!=\!\!\!&0,& 
    \bra{\Ybs{\index}}\qop\ket{\Rbs{\index}}&\!\!\!=\!\!\!&
    \character{\R{}}(q)-2\character{\V{\index}}(q), \vspace{1.5ex}\\
    \bra{\Rbs{\index}}\qop\ket{\Rbs{\index}}&\!\!\!=\!\!\!&\character{\R{}}(q).
  \end{array}  
  \label{eqn:torus-ampl}
\eeqn
Obviously, this does not exactly reproduce the elements of the five-dimensional 
representation given in \eqref{eqn:triplet_characters} but rather linear
combinations of them and the unphysical contribution $\log(q)\Theta_{1,2}(q)$. 
This has to be taken care of when calculating physical relevant boundary 
conditions with the help of Cardy's consistency equation.

\section{With the general method}

The method presented in \cite{BrFl:02} provides an efficient tool for the
investigation of the boundary states under the restrictions of the symplectic 
fermion algebra. It bases on the general ansatz for a boundary state
connecting a holomorphic and an anti-holomorphic representation $\M{h}$ and 
$\Mbar{\ol{h}}$ at the boundary
\beqn
  \ket{\bs}=\sum_{l,m,n} c^{\,l}_{mn}\fuse{\id}{\ol{\U}}\ket{l,m}
  \otimes\ol{\ket{l,n}}.
  \label{eqn:BSansatz}
\eeqn  
The task is to directly calculate the matrix $c$. This is done in an iterative 
procedure. Since the sum in \eqref{eqn:BSansatz} is infinite, the 
coefficients $c^{\,l}_{mn}$ can only be derived up to any finite level $l=L$. 
The idea is that this provides the basis for the second step, the 
identification of the boundary states.

The boundary state consistency equation for this symmetry algebra is given by 
\eqref{eqn:chiB}:
\beqn
  \left(\chi^\pm_m-\e^{\pm\imag\phi}\overline{\chi}{}^\pm_{-m}\right)
  \ket{\bs}=0,
  \label{eqn:chiB1}
\eeqn
where $\phi$ is the spin which can take the values $\phi=0,\pi$ at the boundary,
since we force $\ket{\bs}$ to be compatible with the $\W$-algebra. It is then 
clear that \eqref{eqn:VirB} and \eqref{eqn:WB} are automatically satisfied once 
\eqref{eqn:chiB1} is valid. This implies that the solutions are linear 
combinations of the boundary states of section \ref{section:W-BS}. The 
naturally arising question is, especially when comparing the results presented 
in the two previous sections, whether the fermion symmetry is more restrictive 
than the $\W$-algebra, \ie, if less states are found here than in the latter 
theory. The opposite is the case: Using the method of \cite{BrFl:02} we again
find ten proper boundary states. Denoting the $\phi=0$ case by the quantum 
number $(+)$ and $\phi=\pi$ by $(-)$ as in the previous discussion, these 
states are:
\begin{align}\begin{split}
  \ket{\vac,\vac;\pm}&=
    \ket{\vac,\vac}\pm\ket{\one^+,\one^-}\mp\ket{\one^-,\one^+}+\ldots\,,
    \vspace*{1ex} \\
  \ket{\vac,\logvac;\pm}&=
    \ket{\vac,\logvac}+\ket{\logvac,\vac}
    \pm\ket{\xi^+,\xi^-}\mp\ket{\xi^-,\xi^+}+\ldots\,, \vspace*{1ex} \\
  \ket{\vac,\xi^\spin;\pm}&=\ket{\vac,\xi^\spin}\pm\ket{\xi^\spin,\vac}+
    \ldots\,,~\spin=+,-,\vspace*{1ex} \\
  \ket{\mu,\mu;\pm}&=\ket{\mu,\mu}\pm\ket{\nu^+,\nu^-}\mp\ket{\nu^-,\nu^+}
    +\ldots\,.
\end{split}\label{eqn:sympl-states} \end{align}
Here, $\ket{m,n}$ is used as a short-hand for $\ket{m}\otimes\ol{\ket{n}}$. 
This result may be compared to the one for the $\W$-algebra. We obtain the 
following identities
\beqn
  \begin{array}{ccl@{\qquad}ccl}
    \ket{\vac,\vac;\pm}&\!\!\!=\!\!\!&\ket{\VNbs}\pm\ket{\VObs},&
    \ket{\vac,\logvac;\pm}&\!\!\!=\!\!\!&\left(\ket{\RNbs}+d\ket{\VNbs}\right)
      \pm\left(\ket{\RObs}-t\ket{\VObs}\right), \vspace{1ex} \\
    \ket{\vac,\xi^{\spin};\pm}&\!\!\!=\!\!\!&
      \ket{\RNObs^{\spin}}\pm\ket{\RONbs^{\spin}},&
    \ket{\mu,\mu;\pm}&\!\!\!=\!\!\!&\ket{\VTEbs}\pm\ket{\VOEbs}.
  \end{array}  
\eeqn
This identification uses the fact that the boundary states fulfill
\eqref{eqn:VirB} and \eqref{eqn:WB}. Thus, the first level contributions of 
\eqref{eqn:sympl-states} can be compared to the results of section 
\ref{section:W-BS} to gain the corresponding linear combinations of the states 
given there.

To show that the result \eqref{eqn:KaWh-IBS} of Kawai and Wheater is compatible
to ours one has to keep in mind that the coherent states obey the consistency 
equation \eqref{eqn:chiB1}, and hence \eqref{eqn:VirB} and \eqref{eqn:WB}.
Therefore, they can be expressed in terms of the states
\eqref{eqn:sympl-states}. Indeed, we find
\beqn
  \ket{\bs_{\vac\pm}}=\ket{\vac,\vac;\pm}\mbox{~~and~~}   
  \ket{\bs_{\mu\pm}}=\ket{\mu,\mu;\pm},
\eeqn
up to possible additional contributions from null-states and the different 
normalization. It seems contradictory that here, no boundary 
state based on $\fuse{\logvac}{\ol{\logvac}}$ is found. But reviewing
\cite{KaWh:01} as quoted in section \ref{section:sympl_coherent_states} these 
are the states $\ket{\bs_{\logvac\pm}}$ (or rather $\bra{\bs_{\logvac\pm}}$) 
which occur only as the duals to $\ket{\bs_{\vac\pm}}$ in the Ishibashi states. 
This is remarkable, since in our framework the only states having such 
logarithmic contributions, \ie, $\fuse{\logvac}{\ol{\logvac}}$-like terms, are 
$\ket{\Xbs{\index}}$ that we used in precisely the same manner. This suggests, 
that the coherent states based on $\fuse{\logvac}{\ol{\logvac}}$ are related 
to $\ket{\Xbs{\index}}$ in the same way as above:
\beqn
  \ket{\logvac,\logvac;\pm}=\ket{\XNbs}\pm\ket{\XObs}.
\eeqn
Observe the fact that these states do not exactly correspond to
$\ket{\bs_{\logvac\pm}}$ due to the connection to the local theory as discussed
above. 

The generic procedure of \cite{BrFl:02} yields a much bigger collection 
of states in comparison to Kawai and Wheater. Especially, the mixed boundary
states were not discussed by them and the Ishibashi boundary state for the 
module $\R{}$ was obtained by the identification $2\V{0}+2\V{1}\equiv\R{}$. 
Presumably therefore and by referring to the local theory by setting 
$\fuse{\vac}{\ol{\logvac}}-\fuse{\logvac}{\ol{\vac}}$ to zero, their physical 
boundary conditions differ from the set of Ishibashi states. 

Indeed, we find that the coherent state method produces exactly the same amount 
of states when starting from the ``invariant vacua'' that we have:
\beqn
  \big\{\fuse{\vac}{\ol{\vac}},\,\fuse{\vac}{\ol{\logvac}}
  +\fuse{\logvac}{\ol{\vac}},\,
    \fuse{\vac}{\ol{\xi}^\spin}-\e^{\imag\phi}\fuse{\xi^\spin}{\ol{\vac}},\,
    \fuse{\mu}{\ol{\mu}}\big\}.
\eeqn    

The symplectic fermions decompose the $\Lop_0$ operator in such a way that
\beqn
  \chi^\pm_0\logvac=-\xi^\pm\mbox{~~and~~}\chi^\pm_0\chi^\mp_0\logvac=\mp\vac.
\eeqn
With respect to \eqref{eqn:P} and \eqref{eqn:R01} this suggest that the 
intertwining operators $\Pop$ and $\Pdaggerop$ and the corresponding boundary 
states $\ket{\RNObs^\spin}$ and $\ket{\RONbs^\spin}$ might be closely related 
to the fermionic zero modes.

\section{Discussion}

We worked out the space of boundary states in the rational LCFT with central
charge $\c=-2$ under the restrictions of the symplectic fermion symmetry. It 
turned out that these states coincide with the solution we presented in 
\cite{BrFl:02}. In particular, this implies that the symplectic fermion algebra 
gives no additional constraints on the boundary states in comparison to the
$\W(2,3,3,3)$-algebra of the rational $\c=-2$ LCFT. This is interesting because 
the latter one is embedded in the former. One might guess that the boundary 
state consistency equation for the symplectic fermion symmetry is more 
restrictive than the one for the $\W$-algebra. On the other hand, already in
\cite{BrFl:02} we noticed the close relation between the derivation of boundary 
states and the construction of a local theory (see fig. \ref{fig:R01}). At least
for $\c=-2$ the latter one is uniquely defined which would suggest, that there
exists exactly one consistent solution for the set of boundary states.

To construct the boundary states, we used the same method that we presented in 
\cite{BrFl:02} for the $\W$-algebra case. This shows that this method really
yields a general prescription for the treatment of boundary states and is 
easily adoptable to different frameworks (like the symplectic fermions in this 
case). Thus, it seems natural that the presented results generalize to more 
complicated theories. For the coherent states this was already pointed out by 
Kawai \cite{Ka:021}. 

We compared the results to the coherent state solution of Kawai and Wheater 
and were able to show that both approaches are equivalent, leading to exactly 
the same set of states. However, our results differ in some crucial aspects 
compared to \cite{KaWh:01}: They had to divide out the image of 
$(\Lop_0-\Lbarop_0)$ by hand while in our prescription this is implicitly 
included.

\subsection*{Acknowledgements}

The author wants to thank M.\,R.~Gaberdiel for his remarks leading to this
paper. Many thanks to the participants of the conference about
{\em Non-Unitary/Logarithmic CFT\/}, IHES, 2002 and especially to 
M.\,A.\,I.~Flohr for helpful discussions.

\clearpage

\end{document}